\documentclass{article}

     \PassOptionsToPackage{numbers, compress}{natbib}

     \usepackage[final]{neurips_2024}

\usepackage[utf8]{inputenc} 
\usepackage[T1]{fontenc}    
\usepackage{hyperref}       
\usepackage{url}            
\usepackage{booktabs}       
\usepackage{amsfonts}       
\usepackage{nicefrac}       
\usepackage{microtype}      
\usepackage{xcolor}         
\usepackage{amsmath}        
\usepackage{graphicx}

\title{AI Agent for Education: \\ von Neumann Multi-Agent System Framework\thanks{Jiang, Y.-H., Li, R., Zhou, Y., Qi, C., Hu, H., Wei, Y., Jiang, B., \& Wu, Y. (2024). AI Agent for Education: Von Neumann Multi-Agent System Framework. Conference Proceedings of the 28th Global Chinese Conference on Computers in Education (GCCCE 2024), 77–84. Chongqing, China: Global Chinese Conference on Computers in Education.}}
\author{
Yuan-Hao Jiang$^{1, 2, 3}$, Ruijia Li$^{4}$, Yizhou Zhou$^{5}$, Changyong Qi$^{1}$, Hanglei Hu$^{4}$, \\ \textbf{Yuang Wei}$^{1, 2, 3, }$ \thanks{Corresponding Author: philrain@foxmail.com} \,\, , \textbf{Bo Jiang}$^{1, 3}$, \textbf{Yonghe Wu}$^{4}$\\
\\
$^{1}$~\text{Lab of Artificial Intelligence for Education, East China Normal University}\\
$^{2}$~\text{Shanghai Institute of Artificial Intelligence for Education, East China Normal University}\\
$^{3}$~\text{School of Computer Science and Technology, East China Normal University}\\
$^{4}$~\text{Department of Educational Information Technology, East China Normal University}\\
$^{5}$~\text{School of Design and Engineering, National University of Singapore}
}
\begin{document}
\maketitle
\begin{abstract}
  The development of large language models has ushered in new paradigms for education. This paper centers on the multi-Agent system in education and proposes the von Neumann multi-Agent system framework. It breaks down each AI Agent into four modules: control unit, logic unit, storage unit, and input-output devices, defining four types of operations: task deconstruction, self-reflection, memory processing, and tool invocation. Furthermore, it introduces related technologies such as Chain-of-Thought, Reson+Act, and Multi-Agent Debate associated with these four types of operations. The paper also discusses the ability enhancement cycle of a multi-Agent system for education, including the outer circulation for human learners to promote knowledge construction and the inner circulation for LLM-based-Agents to enhance swarm intelligence. Through collaboration and reflection, the multi-Agent system can better facilitate human learners' learning and enhance their teaching abilities in this process.
\end{abstract}

\section{Introduction}

In recent years, large language models (LLMs) have achieved remarkable success in the field of artificial intelligence (AI) \cite{brown_language_2020, chen_interact_2023}, with their rapid increase in intelligence level being attributed to the effective utilization of comprehensive training datasets and a large number of model parameters. Benefiting from LLM's significant advancements in various downstream tasks such as machine translation \cite{radford_language_2019}, question-answering systems \cite{touvron_llama_2023}, and dialogue systems \cite{touvron_llama_2023}, numerous real-world applications based on LLM have emerged rapidly. By introducing high-quality educational data and educational corpora, LLM's capabilities in educational tasks such as essay assessment, Socratic teaching, and emotional support have been rapidly enhanced \cite{dan_educhat_2023}. With the core ability of self-reflection in LLM, AI Agents can utilize LLM as the brain to empower their environment perception \cite{russell_artificial_2010} and task-solving abilities \cite{radford_language_2019}, effectively facilitating collaborative learning \cite{chen_scalable_2024}, online learning \cite{zivojinovic_application_nodate}, and learning scenario simulations \cite{spoelstra_agent-based_2007} in educational settings.

Thanks to the automated decision-making ability of AI Agents, the system composed of multiple Agents can each perform its duties and solve complex problems through collaboration among AI Agents \cite{chen_generative_2019,radford_language_2019, touvron_llama_2023, touvron_llama_2023-1}. Similar to natural phenomena reflecting collective intelligence such as biological foraging \cite{wei_acdo_2024} and genetic evolution \cite{jiang_control_2023}, multiple intelligent agents can also produce emergent phenomena of "1+1>2" through collaboration among individuals. Through collaboration and interaction, the Multi-Agent System (MAS) can possess extraordinary intelligence and automatically complete complex tasks such as software development, and even learn to deceive other Agents in games to achieve victory \cite{hong_metagpt_2023}. MAS can also replace traditional Intelligent Tutoring Systems (ITS) to better promote learners' knowledge construction process \cite{deng_towards_2023}. Thanks to LLM's rich responsiveness and real-time response capabilities, multiple AI Agents in MAS can concurrently play different roles such as teachers \cite{chi_constructing_1996} and learning companions \cite{simon_usage_2023} to facilitate learners' learning process.

To comprehensively analyze the capabilities of MAS, inspired by the von Neumann Machine structure \cite{zhang_organismic_2020}, this paper proposes the von Neumann MAS Framework (vNMF). vNMF covers four operational stages: task decomposition, self-reflection, memory processing, and tool invocation, and divides MAS into elements such as logic unit, storage unit, control unit, and input-output devices. vNMF also includes a bidirectional cycle model, covering the inner circulation of Agent swarm intelligence emergence and the outer circulation of learner knowledge construction. Based on vNMF, the structural composition and operational logic of MAS can be more easily understood. With vNMF, researchers can better understand how LLM-related technologies, represented by Chain of Thought \cite{wei_chain_2023}, influence the final output of MAS, and teachers or students can better utilize LLM-based-MAS to promote teaching and learning.

In this paper, Chapter 1 introduces the background of the research, Chapter 2 presents the proposed von Neumann MAS framework. Then, Chapter 3 introduces the ability enhancement cycle of MAS for education, including the outer circulation for human learners to promote knowledge construction and the inner circulation for LLM-based Agents to enhance swarm intelligence. Finally, Chapter 4 summarizes the work of this paper.

\section{von Neumann MAS Framework}

\begin{figure}[t]
  \centering
  \includegraphics[width=\linewidth]{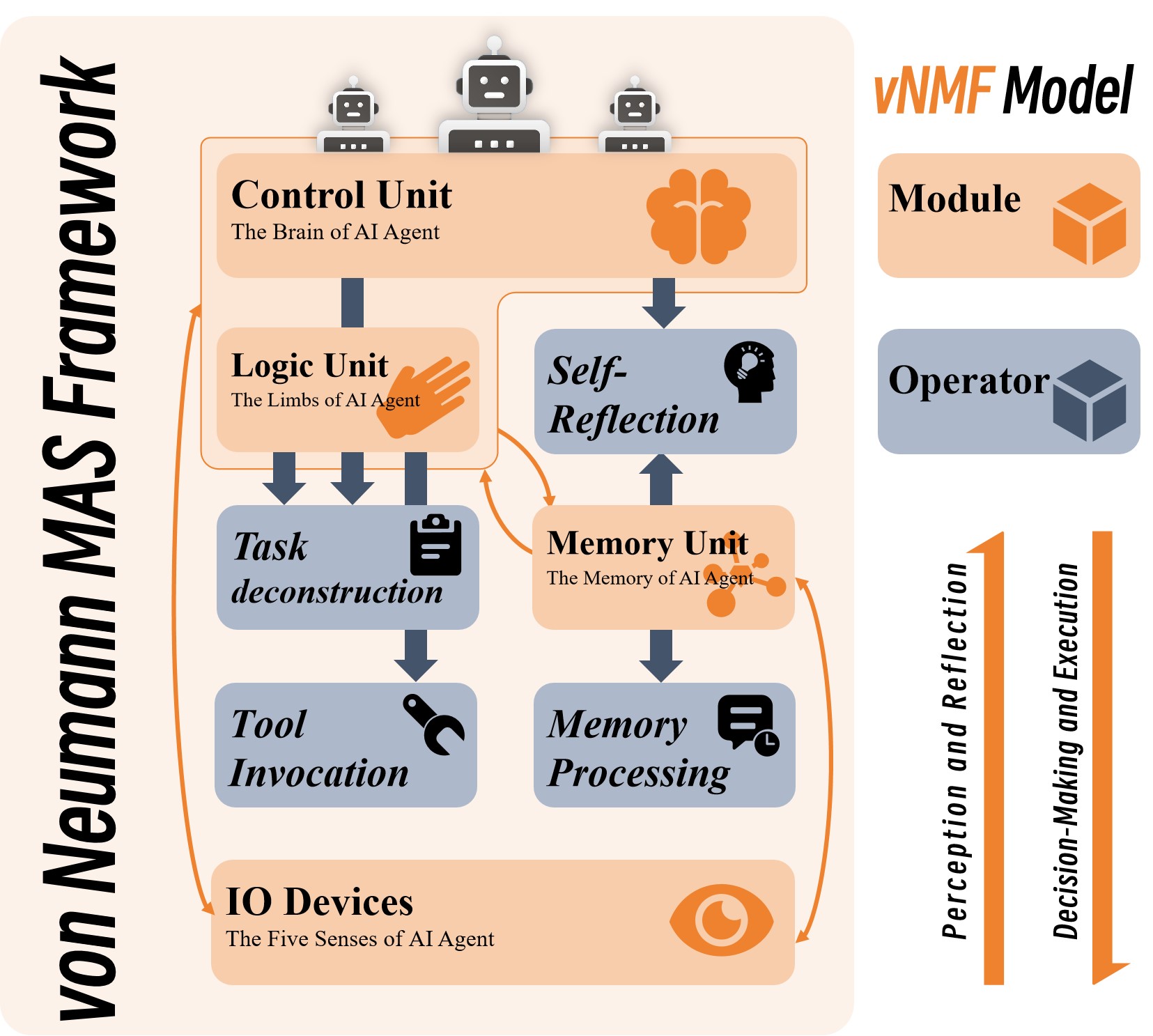}
  \caption{von Neumann multi-Agent system framework.}
  \label{fig:1}
\end{figure}

Upon deliberation of information flow, the architecture of LLM-based AI Agents is delineated into four core components: planning, action, tools, and memory \cite{openai_gpt-4_2023}. AI Agents engage in problem analysis through the planning module, amalgamating historical insights with the memory module to foster cogitation, harnessing the tools module to assimilate external resources, and ultimately operationalizing thought outcomes via the action module. For a comprehensive comprehension of AI Agent constituents, further classification into perception, brain, and action is proposed \cite{xi_rise_2023}. Through the perception module, AI Agents procure environmental data, leveraging LLM to enhance cognitive and planning proficiencies, culminating in environmental influence via the action module. In contrast to prevailing frameworks, we introduce the vNMF model, drawing inspiration from the von Neumann Machine's architecture, to interlink AI Agent components, as depicted in Figure \ref{fig:1}. AI Agent operations are delineated into four categories: task decomposition, self-reflection, memory processing, and tool invocation. Furthermore, AI Agents are deconstructed into four modules: logic unit, memory unit, control unit, and input-output (IO) devices. Notably, the logic unit and control unit jointly comprise the AI Agent's "Central Processing Unit," serving as the bedrock for enhancing the Agent's self-capabilities.

In the proposed vNMF model, the logic unit operates akin to the limbs of the AI Agent, endowed with the capability to activate external tools and execute specific tasks. The memory unit retains the memories of AI Agents, facilitating the recollection of past experiences to refine their actions. Serving as the brain of the LLM-based-Agent, the control unit orchestrates the coordination among multiple AI Agent modules. Functioning as the core component of the AI Agent, the control unit collaborates with the logic unit for task decomposition and with the memory unit for self-reflection. 

Furthermore, the input-output devices are tasked with procuring external environmental data for the AI Agent and disseminating thinking or execution outcomes. This module encompasses components such as a Graphical User Interface (GUI), multimodal sensors, or external robotic entities. Through the synergy of these four modules in task decomposition, self-reflection, memory processing, and tool invocation, the individual AI Agents' capabilities can be enhanced, leading to improved performance of the MAS particularly in educational settings. Subsequently, leveraging the proposed vNMF model, we will expound on the perspectives of task decomposition, self-reflection, memory processing, and tool invocation.

\subsection{Task Deconstruction}

Through task decomposition, AI Agents can partition large tasks into smaller, more manageable sub-goals, thereby enhancing their efficiency in handling complex tasks. For instance, when tackling intricate mathematical problems, AI Agents employ task decomposition to segment them into multiple steps, resulting in heightened accuracy and clearer presentation of the thought process to learners. A prominent method of task decomposition, Chain of Thought (CoT) \cite{wei_chain_2023}, has evolved into a widely embraced Prompting technology in the industry, significantly enhancing model performance in complex tasks. By instilling a "think step by step" approach in the model, CoT dissects intricate tasks into simpler subtasks, elucidating the model's thought process with greater clarity and precision.

To further refine CoT's performance, Tree of Thoughts (ToT) \cite{yao_tree_2023} is proposed to circumvent potential erroneous paths during problem decomposition. ToT initially dissects the problem into logically rigorous steps, with each step corresponding to a distinct thought, thereby forming a tree-like structure of thoughts where each node represents a possible direction of thought. During the exploration process, ToT can adopt breadth-first search (BFS) or depth-first search (DFS) strategies. The BFS strategy systematically explores the first-level child nodes of the current node before proceeding to the subsequent level, while the DFS strategy delves deeply into each branch, exhaustively exploring all possible paths. This extension of CoT by ToT enables comprehensive and in-depth problem-solving, yielding more accurate and comprehensive solutions. Moreover, by incorporating aggregation, backtracking, loops, and other intricate operations, Graph of Thoughts (GoT) \cite{besta_graph_2023} is introduced, concurrently supporting multi-chain, tree-like, and arbitrary graph-shaped Prompt schemes, thereby equipping AI Agents with the capability to tackle complex problems.

In addition to adopting CoT, ToT, GoT, and other LLM-based task decomposition methods, AI Agents can harness external classical Planners for broader sequence overall planning, a technique termed LLM+Planner (LLM+P) \cite{liu_llmp_2023}. LLM+P utilizes planning domain definition language (PDDL) as an intermediary interface for describing planning problems. Initially, LLM translates the problem into "problem PDDL," subsequently soliciting the classical Planner to generate a PDDL plan based on the existing "domain PDDL," and ultimately translating the PDDL plan back into natural language. Through LLM+P, AI Agents can effectively leverage existing problem planners to enhance their task decomposition proficiency in specific domains, thereby averting incorrect paths or insoluble loops.

\subsection{Self-Reflection}

In the instructional setting, errors or excessively intricate content might yield counterproductive outcomes, exacerbating confusion among learners. Through self-reflection, AI Agents can engage in self-critique and retrospection of past actions, assimilate lessons from errors, and refine future actions to elevate the quality of the ultimate output. To augment the self-reflective capacity of AI Agents, Reson+Act (ReAct) \cite{yao_react_2023} integrates reasoning and action within LLM, facilitating more effective interaction of AI Agents with their environment. ReAct delineates clear cognitive steps, guiding AI Agents to structure their thinking across three phases: thought, action, and observation. Grounded in ReAct, AI Agents can achieve heightened performance in tasks demanding a substantial degree of knowledge or decision-making prowess \cite{yao_react_2023}.

In contrast to the ReAct, Reflexion \cite{shinn_reflexion_2023} presents a framework rooted in the principles of reinforcement learning, to enhance AI Agents' reasoning capabilities through the endowment of dynamic memory and self-reflection capacities. This framework adheres to standard reinforcement learning (RL) protocols, where the reward model furnishes simplistic binary rewards, and the action repertoire is modeled after ReAct, introducing linguistic elements into the action domain of specific tasks to execute intricate reasoning steps. Upon detecting suboptimal planning by AI Agents, Reflexion resets the state of the AI Agent and amalgamates unsuccessful trajectories with ideal reflections to guide subsequent planning iterations, leveraging historical experiential lessons to foster reflection and glean insights from failure.

To emulate the multi-threaded reasoning processes inherent in multi-agent collaborations, multi-Agent debate (MAD) \cite{du_improving_2023} is proposed. MAD endeavors to mitigate illusions or misconceptions through iterative debates among multiple Agents. In each debating round, each AI Agent must contemplate not only its response but also responses from other AI Agents, crafting new responses based on this collective input. Within this framework, each agent assumes the dual roles of validator and improver, tasked with generating novel responses predicated on the received inputs. This iterative process can be iterated across multiple rounds to incentivize AI Agents to reflect on their conduct and enhance their ultimate performance.

\subsection{Memory Processing}

Through the synthesis and retrospection of memory, AI Agents can manage information of longer duration and greater complexity. Within the educational realm, interrogating memory allows for the analysis of a learner's current grasp of knowledge, thereby facilitating the provision of higher-quality feedback to enhance teaching efficacy. Memory can be classified into short-term memory and long-term memory based on its temporal generation. Short-term memory retains information that is presently within our awareness, whereas long-term memory has the capacity for prolonged information storage. Short-term memory functions in the context of learning, constrained by the context window length of the transformer structure, while long-term memory serves as an external vector storage that agents can attend to during retrieval. Long-term memory is further subdivided into declarative memory, housing facts and events, and procedural memory, which encompasses unconscious skills that can be executed automatically. MetaGPT presents an effective memory processing framework aimed at facilitating the rapid response of MAS to high-concurrency requests \cite{hong_metagpt_2023}. By adeptly processing and organizing the information embedded within memory, multiple AI Agents within MAS can better preserve their distinct individual traits and demonstrate enhanced anthropomorphism within classroom simulation environments \cite{jinxin_cgmi_2023}.

\subsection{Tool Invocation}

Utilizing tools represents a significant trait of high-level human intelligence. AI Agents acquire the ability to access external application APIs, with the external tools at their disposal significantly augmenting the model's functionality. Proficiency in invoking external tools enables AI Agents to enhance the teaching process. For instance, they can initiate the analysis of input question images using computer vision models, employ optical character recognition techniques to derive text representations from educational resources and employ external computing tools for solving calculus or other intricate operations. HuggingFace stands out as an open-source platform boasting a plethora of models accessible for AI Agents. Leveraging the HuggingFace platform, HuggingGPT \cite{shen_hugginggpt_2023} serves as a framework employing ChatGPT as a task planner to select models based on descriptions and succinctly summarize responses derived from execution outcomes. Furthermore, through Tool Augmented Language Models (TALM), AI Agents can adeptly identify suitable tools \cite{parisi_talm_2022}. By enabling AI Agents to acquire proficiency in utilizing existing tools, their operational scope is broadened, allowing them to execute more intricate operations.

\section{The Ability Enhancement Cycle of MAS for Education}

\begin{figure}[t]
  \centering
  \includegraphics[width=\linewidth]{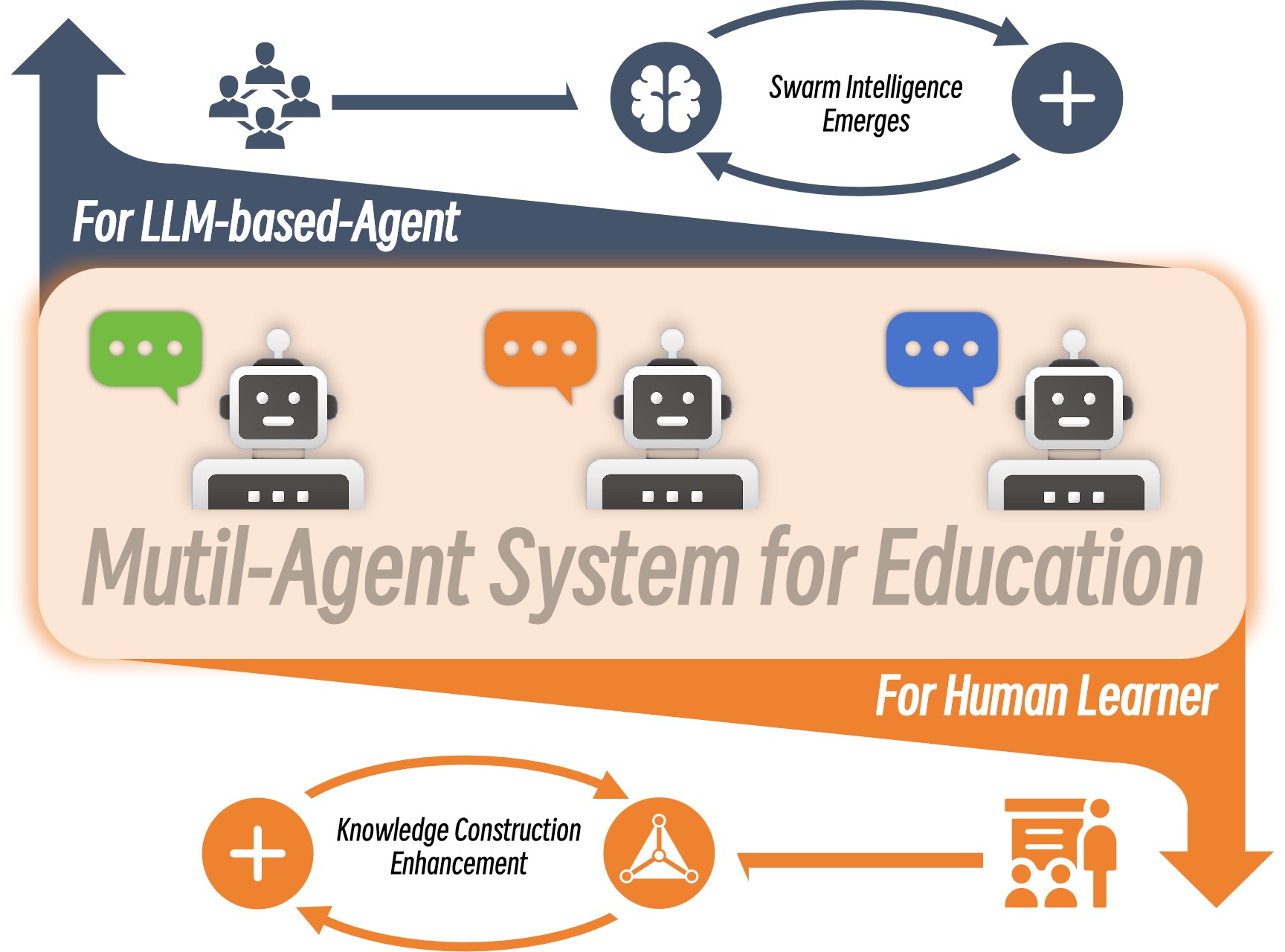}
  \caption{The ability enhancement cycle of MAS for education.}
  \label{fig:2}
\end{figure}

In the preceding sections, we have delved into the structure of MAS. To delve into how MAS fosters the learning journey of students in educational settings and continually evolves within this context, we introduce an ability enhancement cycle model tailored for MAS in education. This cycle encompasses both inner circulation for LLM-based-Agents and outer circulation for human learners, as depicted in Figure \ref{fig:2}.

\subsection{Outer Circulation for Human Learners: Knowledge Construction Enhancement}

Within the outer loop of MAS, human learners enrich their learning journey through human-computer collaboration, aiming for effective knowledge acquisition. MAS operates as a unified entity, refining output outcomes and bolstering pedagogical capacities through collaborative efforts among multiple intelligent agents \cite{du_improving_2023}. Moreover, it facilitates the simultaneous adoption of various roles by multiple AI Agents, including educators, learning companions, and more, to streamline the learning process \cite{jinxin_cgmi_2023}. As learning companions, AI Agents contribute to enhancing the learning experience. Leveraging artificial intelligence technology, they furnish learners with tailored learning experiences and educational resources by delivering timely support, explanations, inquiries, and content. They also undertake metacognitive scaffolding responsibilities, intervening promptly in learners' educational journeys \cite{neira-maldonado_intelligent_2024}. Through the integration of teaching resources grounded in extensive data analysis and real-time access to network resources facilitated by AI Agents, the system amalgamates diverse educational materials and courseware. This provision of abundant teaching resources and case studies aims to guide learners through online learning effectively, akin to instructors \cite{wei_study_2021}. This approach fosters the knowledge construction process among learners and enhances learning outcomes.

\subsection{Inner Circulation for LLM-based-Agent: Swarm Intelligence Emerges}

As the outer loop of MAS operates, it enriches the knowledge construction process among human learners. Concurrently, within the inner loop of MAS, it fosters the emergence of swarm intelligence among LLM-based Agents. In the natural realm, the macroscopic intelligent behavior displayed by social organisms through collaboration is termed swarm intelligence \cite{tang_adaptive_2023, zhou_face_2023}. Through effective collaboration among multiple AI Agents, MAS's capabilities are enhanced, facilitating the adept handling of intricate tasks such as software development \cite{hong_metagpt_2023}. Educators can simulate and forecast collaborative outcomes among learners at varying proficiency levels through collaboration and simulation among multiple Agents, thus providing a foundation for differentiated instruction \cite{spoelstra_agent-based_2007}. Additionally, multiple AI Agents can refine output structures through debate. Despite the potential for erroneous information output by LLM, the stochastic nature of these errors is likely to be mitigated in this cross-validation process \cite{du_improving_2023}. Through the effective promotion and utilization of swarm intelligence in MAS, the teaching capabilities of MAS are significantly augmented. Furthermore, MAS can leverage interaction data for model fine-tuning, continually enhancing output results throughout the teaching process \cite{casper_open_2023}.

\section{Conclusion}

This paper introduces the von Neumann MAS framework and delineates the ability enhancement cycle of MAS for education based on this framework. We posit that MAS can realize both the outer circulation for human learners, fostering knowledge construction, and the inner circulation for augmenting Swarm Intelligence among LLM-based Agents. By bolstering teaching capabilities, MAS can better assume the roles and responsibilities of learning companions, educators, or educational tools for learners, aiding them in effectively constructing knowledge.

\section*{Acknowledgements}

This work was partially supported by the National Natural Science Foundation of China under Grant 61977058, and the Natural Science Foundation of Shanghai under Grant 23ZR1418500. The authors would also like to thank Zi-Wei Chen for her assistance with the document preparation process of this paper.

\bibliographystyle{IEEEtran}
\bibliography{sample-base}

\end{document}